\documentclass[%
aip,
jmp,%
amsmath,amssymb,groupedaddress,
reprint,%
]{revtex4-1}
\usepackage{hyperref} %adds links
\usepackage{graphicx}% Include figure files
\usepackage{dcolumn}% Align table columns on decimal point
\usepackage{bm}% bold math
\newcommand{\eq}[1]{\begin{equation}#1\end{equation}} % for equation,
\usepackage[mathlines]{lineno}% Enable numbering of text and display math
\begin{document}
	
	%\preprint{AIP/123-QED}
	
	\title[Fast thermometry with a hysteretic proximity Josephson junction]{Fast thermometry with a proximity Josephson junction}% Force line breaks with \\
	%\thanks{Footnote to title of article.}
	
	\author{L. B. Wang}
	\email{libin.wang@aalto.fi.}
	\author{O.-P. Saira}
	\author{J. P. Pekola}
	\affiliation{Low Temperature Laboratory, Deparement of Applied Physics, Aalto University, P.O.Box 13500, FI-00076 Aalto, Finland.}%
	
	\date{\today}% It is always \today, today,
	
	\begin{abstract}
		We couple a proximity Josephson junction to a Joule-heated normal metal film and measure its electron temperature under steady state and nonequilibrium conditions. With a timed sequence of heating and temperature probing pulses, we are able to monitor its electron temperature in nonequilibrium with effectively zero back-action from the temperature measurement in the form of additional dissipation or thermal conductance. The experiments demonstrate the possibility of using a fast proximity Josephson junction thermometer for studying thermal transport in mesoscopic systems and for calorimetry.
	\end{abstract}
	
	\maketitle
	
	Thermometry is a cornerstone in studies of thermodynamics. When the investigated system is in equilibrium, the working speed of a thermometer may not be an important factor, as the system status does not change with time. In the past decades, much progress has been made in understanding thermal transport in nanoscale systems in the steady state \cite{Berut2012,Schwab2000,Meschke2006}. If the time scale of interest is shorter than thermal relaxation time $\tau$ of the relevant system, one needs to measure the system temperature with a fast thermometer in nonequilibrium. The relaxation time increases with lowering temperature, which makes the thermal relaxation time of electrons experimentally accessible at millikelvin temperatures. A thermometer with large bandwidth is still needed to expand the temperature range and the variety of processes that can be observed in non-equilibrium.

	Fast thermometry with sub-$\mu$s time resolution has been realized with Normal Metal-Insulator-Superconductor (NIS) tunnel junctions and superconducting weak links embedded in resonant circuits\cite{Schmidt2003,Schmidt2005,Gasparinetti2015,Saira2016,Govenius2016}. In these methods, the measurement bandwidth is set by the linewidth of the resonant circuit, which can not be increased indefinitely without sacrificing the readout sensitivity. Recently, Ref.$\left[ 9\right]$  has shown nanosecond thermometry using a superconducting nanobridge, introducing the hysteretic JJ as a fast thermometer for calorimetry with easy integration. 

	In this letter, we perform fast, minimally invasive thermometry of an evaporated thin-film using proximity JJs. Instead of using superconducting nanobridge, we utilize proximity JJs consisting of a normal metal weak link contacting two superconducting leads. The normal section of the weak link is galvanically connected to the thin film under study, whose electron temperature can be elevated by Joule heating pulses. We devise a probing scheme that allows us to study the nonequilibrium electron temperature in the thin film with ${\mu}$s time resolution, vanishing dissipation, and virtually zero added heat conductance to the system under study prior to the measurement pulse. Experimental results show the great potential of using proximity JJ thermometer for precision and fast measurements of electron temperature in metallic films.
	\begin{figure}[]
	\centering
	\includegraphics[width=0.8\textwidth]{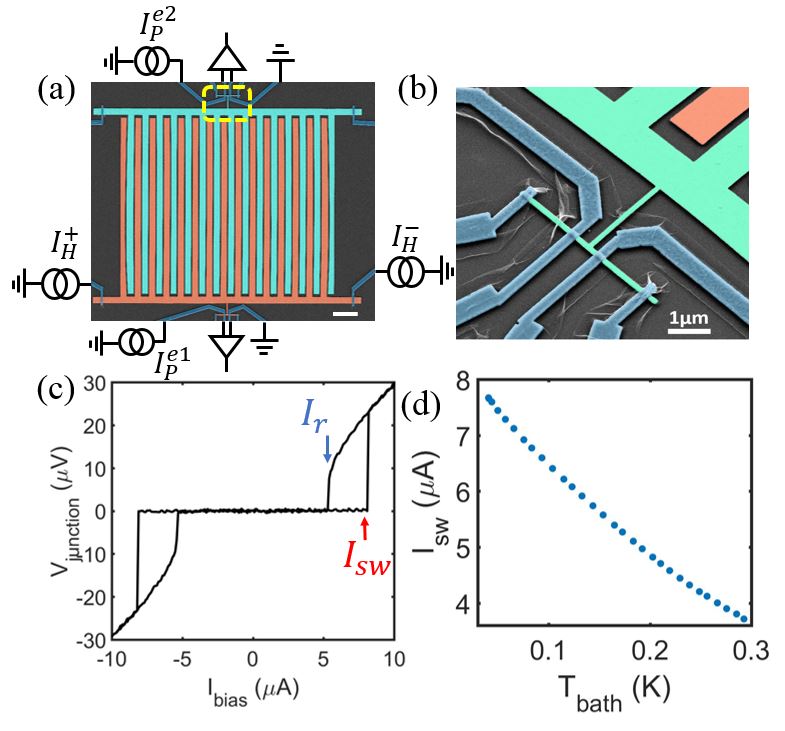}
	\caption{(a) False-color SEM images of a device and its measurement circuit. Normal metal (cyan and pink), Al (blue), scalebar 4~$\mu$m. (b) Zoom of the dashed yellow area in (a) shows the JJ thermometer. (c) I-V curve of the thermometer shows hysteresis at 60~mK. (d) Temperature dependence of $I_{sw}$, Based on this dependence, SNS JJ works as an electron thermometer.}
	\label{thermometer}
\end{figure}

	The JJ thermometer consists of a normal metal wire (cyan) sandwiched between Al superconducting electrodes (blue), shown in Scanning Electron Microscope (SEM) image in Fig.~\ref{thermometer}(b). The inner four electrodes are used in the experiment to measure the switching current of the JJ, the other two Al electrodes at the end of the wire are used to determine the contact resistance between Al and normal metal. The thermometer is connected to interdigital normal metal films by a narrow metal wire. In Fig.~\ref{thermometer}(a) we show the whole device structure together with the measurement circuit. Two interdigital normal metal films (pink and cyan) form thermal coupling to the local substrate without a galvanic connection. Electrons in one of the films (pink) are Joule heated by applying current through Al electrodes to the normal metal heater. Currents with different polarities ($I_H^+$, $I_H^-$) are applied to the heater contacts to ensure no heating current flows through the weak link. The length of the normal metal heater is 42 $\mu$m with the resistance of $8.5~\Omega$. The calculated electron diffusion time $\tau_D = \frac{L^2}{D}$ in the heater is around 40 ns, where $L$ is junction length and $D$ = 140 cm$^2$/s is the diffusion constant. Typically for metals at low temperature, electron--electron relaxation time is around 1 ns, which is much smaller than diffusion time in the metal film, indicating a well-defined Fermi distribution in the films. Within the temperature range studied in this experiment, the wavelength of the thermal phonons in metal films and in the substrate is typically on the order of micrometre, then the two metal films with a distance of 200~nm from each other have the same phonon temperature as the local substrate.

	The devices are fabricated on a silicon wafer coated with 300~nm silicon dioxide. Two-step e-beam lithography is used to define normal metal films and superconducting electrodes separately. Metal films with a thickness of 50~nm are first deposited by e-beam evaporation. Before contacting normal metal with Al, argon plasma cleaning is used to remove the residual resist from the surface of the normal metal, followed by deposition of 3~nm of Ti between Al and metal film to ensure a good contact. Devices are cooled down with homemade plastic dilution refrigerator. All measurement lines are filtered with on-chip RC filters at the temperature of the mixing chamber of the refrigerator.

	The switching process of a JJ is known to exhibit stochastic character due to thermal and quantum fluctuations\cite{Martinis1987}. In the case of DC measurement, by ramping up biasing current through JJ, one can drive the JJ from superconducting state to the resistive state as shown in Fig.~\ref{thermometer}(c). Here, the junction shows a $I_{sw}$ of 7.8~$\mu$A with normal state resistance of 3~$\Omega$, the switching current $I_{sw}$ is defined as the corresponding biasing current when JJ switches to the resistive state. When sweeping back the biasing current, the junction switches from resistive state to superconducting state at a biasing current well below $I_{sw}$. This suppression of the retrapping current ($I_r$) originates from the overheating of the electrons in metal wires after junction switches to the resistive state. For the measured device, thermal hysteresis is observed at temperatures up to 250~mK. Importantly, the dissipation begins only after the switch to the normal state. Hence, the statistics of the switching current provide information about the unperturbed film temperature. For all the junctions measured, the calculated Thouless energy $\epsilon_{c}$ is about 30~$\mu$eV, much smaller than the superconductor gap of Al ($\Delta \approx$ 200~$\mu$eV), indicating that the system is in the long junction limit\cite{Dubos2001}. Temperature calibration of the JJ thermometer is obtained by varying the bath temperature of the refrigerator and recording the $I_{sw}$, as shown in Fig.~\ref{thermometer}(d). $I_{sw}$ depends almost linearly on temperature without saturation down to 60~mK. With this calibration, the SNS JJ serves as an electron thermometer. 
	
	In steady state, considering a system with constant heating $\dot{Q}_H$ applied to it, the change of system temperature can be expressed as
	
	\eq{
		\Delta T_e = \frac{\dot{Q}_H}{G_{th}}. \label{eqsteadystate}
	}
	Here $\Delta T_e = T_e - T_p$, and $G_{th}$ is the thermal conductance from system to its environment. For normal metal, it is well known that electrons (e) are decoupled from the phonon (p) environment at low temperatures, thermal conductance between electrons and phonons ($G_{e-p}$) is the bottleneck for energy dissipation at low temperature\cite{Little1959,Gantmakher1974,wellstood94}, which leads to the hot electron effect when $G_{e-p}\ll G_K$. Here $G_K$ characterises the phonon--mediated heat transport from the metal lattice to the substrate. Theories and experiments show that the energy flow rate $P_{e-p}$ from the electron gas at temperature $T_{e}$ to the phonon gas at $T_{p}$ is $P_{e-p}= {\Sigma}V(T_{e}^{n}-T_{p}^{n})$. Here $V$ is the metal volume, the exponent $n$ and the material specific e-p coupling constatn $\Sigma$ will be discussed later in details. The thermal conductance between electrons and phonons is $G_{e-p}=n{\Sigma}VT_{e}^{n-1}$ for small temperature differences, i.e. when $T_{e} \approx T_{p}$. 
	\begin{figure}[]
		\centering
		\includegraphics[width=0.8\textwidth]{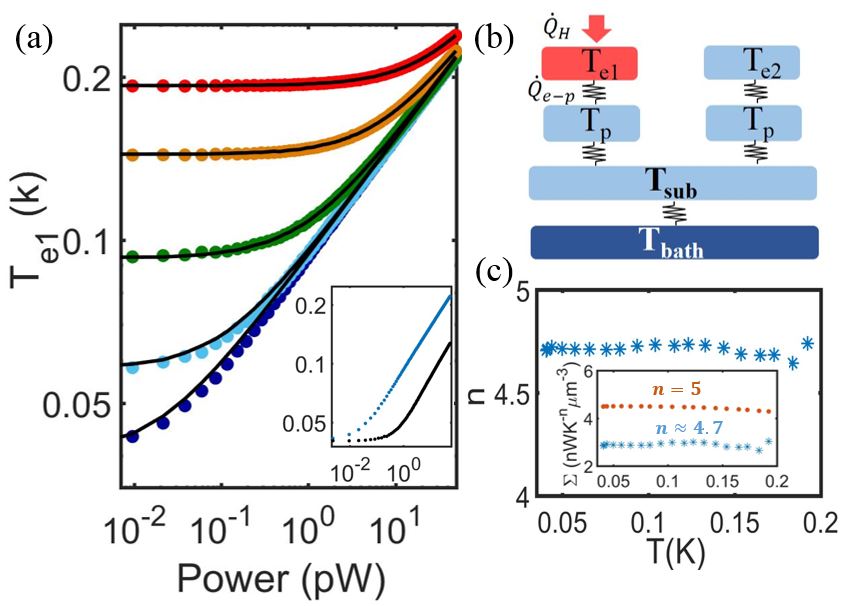}
		\caption{(a) Electron temperature of the heated film as a function of the heating power applied. Black lines are fits with Eq.~(\ref{epcoupling}) with  $T_{bath}$ = 41, 57, 93, 144 and 192~mK from blue to red. Inset: $T_{e1}$ and $T_{e2}$ against heating power at bath temperature of 41~mK. Units are same as in the main plots.(b) The thermal model of heat flows in the system. (c) Temperature dependence of exponent $n$ obtained by fitting of measurements to Eq.~(\ref{epcoupling}) with $n$ and $\Sigma$ as free parameters. Inset: $\Sigma$ as the function of temperature of an Ag film with thickness of 50~nm, fitted by fixing $n$ to 5 (yellow stars) and 4.7 (blue dots).}
		\label{steadystate}
	\end{figure}
	
	We utilize the hot electron effect under steady state conditions to measure the e-p coupling constant in normal metal.  Electron temperature in the metal film is elevated by applying constant heating $\dot Q_H$ on it while measuring its electron temperature ($T_{e1}$) and also the electron temperature of the indirectly heated metal film ($T_{e2}$). Figure~\ref{steadystate}(b) shows the heat flow in the system. Joule heating ($\dot Q_H$) applied to the metal film heats up its electrons, the electrons are coupled to phonons with energy flow rate $\dot Q_{e-p}$. Lattice phonon temperature of the two films ($T_p$) is kept constant at substrate phonon temperature ($T_{sub}$). Here we consider the Kapitza resistance is negligible compared to the thermal resistance between electrons and phonons. Meanwhile, $T_{e2} = T_{p}$ because there is no energy flow between electrons and phonons in steady state in the indirectly heated film. The substrate temperature ($T_{sub}$) near metal films may show a higher temperature than the bath temperature of the refrigerator as Joule heating applied on the metal film heats up the local substrate phonons as well. Electron temperature of the heated metal film as a function of the heating power applied is shown in Fig.~\ref{steadystate}(a). In the inset of Fig.~\ref{steadystate}(a) we show the increased substrate temperature with respect to the bath temperature of the refrigerator at the bath temperature of 41~mK.
	
	With temperature below about 300~mK, heat flow through Al contacts is negligible because of the good thermal isolation of superconducting Al\cite{Peltonen2010}. Then at steady state, the dominant mechanism for electrons in the heated film to cool is e-p scattering. So in steady states, we have
	\eq{
		I^2R = \Sigma V(T_{e1}^n - T_{e2}^n). \label{epcoupling}
	}
	Fitting measurement results to Eq.~(\ref{epcoupling}) with $\Sigma$ and $n$ as free parameters at each bath temperature point, we get $n \approx 4.7$ for the Ag film between 40--200~mK, as shown in Fig.~\ref{steadystate}(c).
	
	The exponent $n$ in Eq.~(\ref{epcoupling}) was first measured by Roukes et al\cite{Roukes1985} for pure Cu films and experiments show that the exponent $n$ equals 5, which can be explained by a theory based on a clean three-dimensional free-electron model with $ql\gg1$. Here, $q$ is the phonon wave factor, and $l$ is the electron mean free path. Phonon wave factor $q$ can be further expressed as $q = hv_S/k_BT$, where $v_S$ is the speed of sound of the phonon mode in metal, $k_B$ is the Boltzmann constant and $h$ is the Planck constant. In a dirty limit with $ql<1$, theories predict that $n$ ranges from 4 to 6 depending on type and level of disorder\cite{Rammer1986,Sergeev2000}. Deviations from $n = 5$ are experimentally observed in normal metals and alloys\cite{DiTusa1992,Hsu1999,Gershenson2001,Karvonen2005}. But this hypothesis is still not fully verified as some experiments show that even for samples in the dirty limit, the energy flow rate from electrons to phonons follows the $T^5$ dependence\cite{Vinante2007,Echternach1992}. The effect of phonon dimensionality and substrate properties on e-p coupling have also been discussed \cite{Underwood2011,Cojocaru2016,Karvonen2007,DiTusa1992}. Experiments show that $n$ falls below 4.5 for phonons in two dimensions\cite{Karvonen2007}. Though there are some discussions about the exponent, $n = 5$ is still the mostly observed dependence in metal films and it is widely used in modelling e-p coupling problems. In our experiments, the low temperature resistivity of the films is $1\times10^{-8}$ $\Omega$$\cdot$m, which gives a mean free path of 81.5~nm. In the temperature range of 60-200~mK, $ql \approx 0.2-1.3$.

	By fitting measurement results to Eq.~(\ref{epcoupling}) with fixed exponent $n = 5$, we get the temperature dependence of the e-p coupling constant $\Sigma$ of Ag shown in the inset of Fig.~\ref{steadystate}(c). In the same plot, we also show the results by fitting experimental data with $n \approx 4.7$. $\Sigma$ shows essentially no temperature dependence as expected with a constant value of about 4.5~nWK$^{-5}$$\mu$m$^{-3}$, compatible with earlier measurements on Ag films\cite{Steinbach1996}, proving the accuracy of our JJ thermometer.
	
\begin{figure}[]
	\centering
	\includegraphics[width=0.8\textwidth]{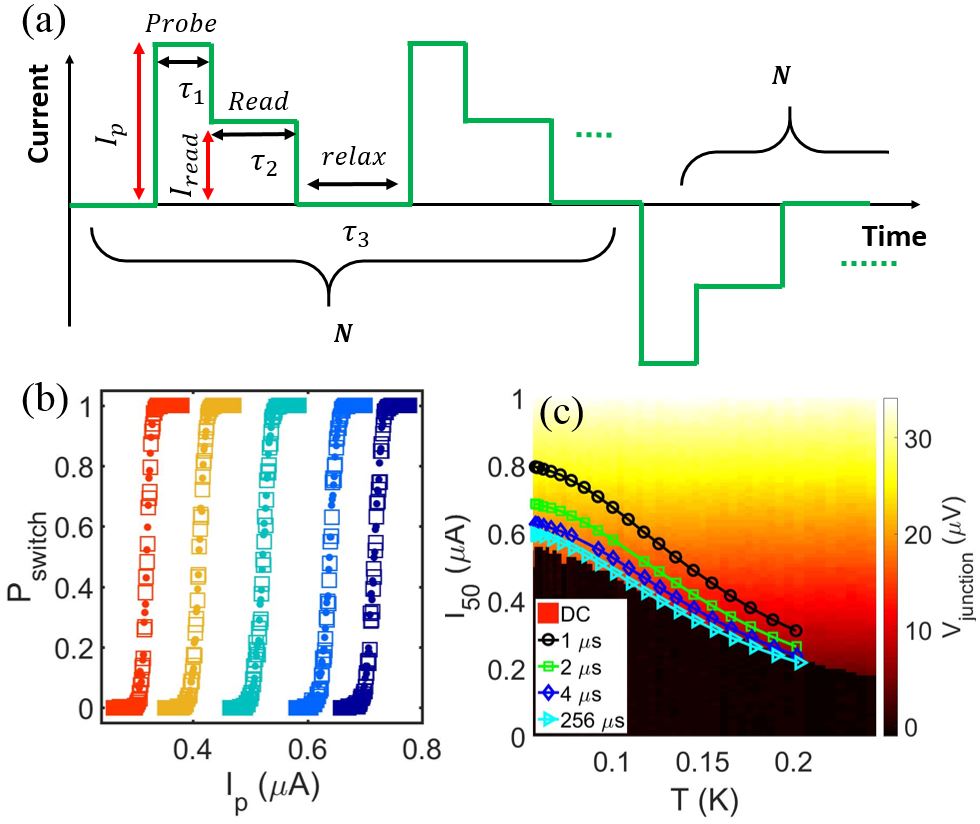}
	\caption{(a) The waveform of current pulse train used in determining $I_{sw}$. (b) Switching probability as a function of probe pulse amplitude at different bath temperatures. $T_{bath} = 60, 90, 121, 151$ and 182~mK from red to blue. (c) Temperature dependence of $I_{sw}$ with varying probe pulse width. DC measurement results are shown in the colour plot.}
	\label{switching}
\end{figure}
	
	The switching process of JJ can also be probed by sending a current pulse to the junction and measuring its response. The probability of the junction to switch to the normal state depends on amplitude and width of the current pulse sent to JJ due to the stochastic character of the switching process\cite{Foltyn2015}. During measurement, a series ($N$) of rectangular current pulses, shown in Fig.~\ref{switching}(a), are sent to the junction and its response is recorded. Each current pulse consists of two parts: probe pulse and read-out pulse. Read-out amplitude $I_{read}$ is kept at a level just above the retrapping current $I_r$ for recording of the switching events. Probe pulse amplitude $I_p$ is varied to probe the switching events. For a particular pulse amplitude, the number of switching events ($n$) is counted, and the switching probability is defined as $P = n/N$. Current pulses with different polarities are used to further check the consistency of the measurement method. The time interval between two current pulses is set to $\tau_3 = 10$~ms in order to ensure cooling of the electrons after the JJ retraps to the superconducting state.

	Figure.~\ref{switching}(b) shows the switching probability against the probe pulse amplitude at different bath temperatures for another sample. Probe pulse width $\tau_1$ and the read-out pulse width $\tau_2$ are set to 2~$\mu$s and 1~ms. Current pulses with different polarities are plotted as dots and squares separately and the results overlap as expected. The switching probability increases from 0 to 1 when increasing the probe pulse amplitude $I_p$. For thermometer calibration, we define $I_{50}$ as that corresponding to the switching probability $P = 0.5$. In Fig.~\ref{switching}(c), we plot the temperature dependence of $I_{50}$ for different probe pulse widths together with quasi-DC sweeps of the bias current. For a fixed bath temperature, longer pulse width gives higher switching probability of the JJ. Thus, smaller $I_p$ is needed to drive JJ to the normal state\cite{Foltyn2015}, so that the measured $I_{50}$ is lower. For pulse width of $256~\mu$s, $I_{50}$ is nearly equal to $I_{sw}$ obtained from the DC measurement.

		\begin{figure}[]
		\centering
		\includegraphics[width=0.8\textwidth]{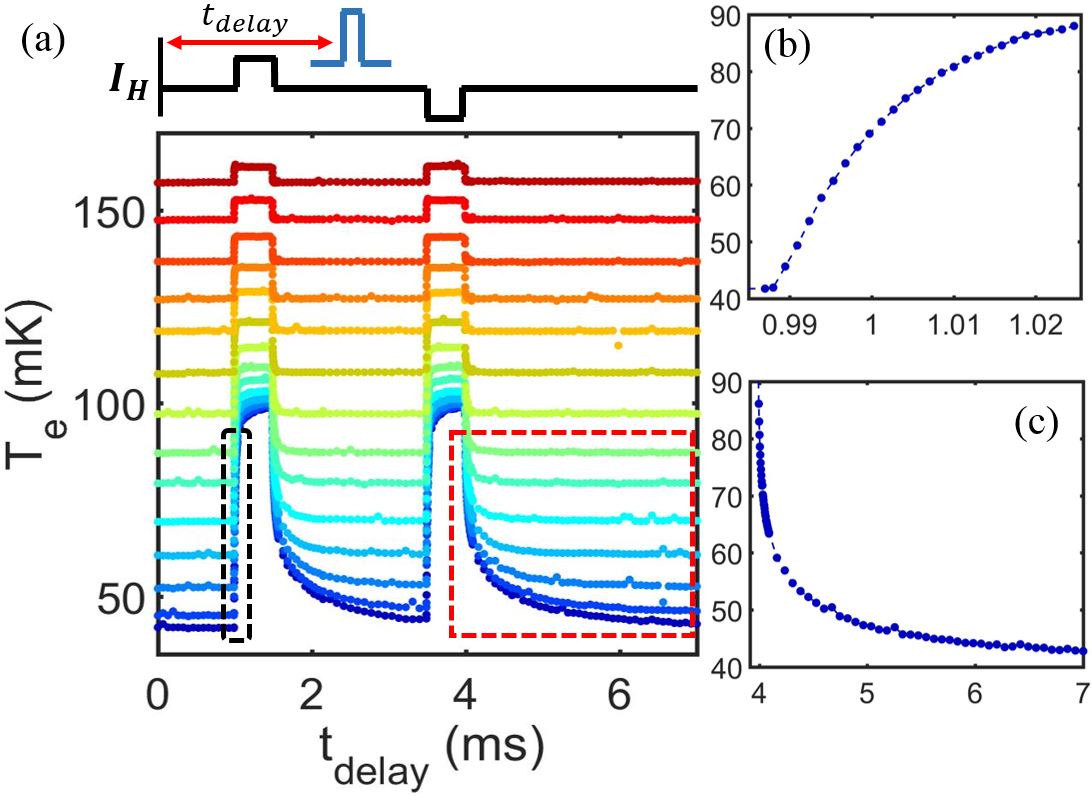}
		\caption{(a) Electron temperature of metal films in response to a heating pulse with bath temperature varying from 41 to 157~mK. $t_{delay}$ is defined as the time interval between heat pulse (black) and probe pulse (blue). (b) and (c) are zoom-in of  black and red squares (a) with the bath temperature of 41~mK.}
		\label{relaxationmeasurement}
	\end{figure}
	Instead of using DC current to elevate the electron temperature in metal film in the steady state, we heat the electrons in the metal film with a rectangular current pulse. As in steady state measurement, pulses with different polarities are used to check that no heating current flows through the probing lines. Meanwhile, by changing the time interval ($t_{delay}$) between probe pulse and heat pulse with one of them fixed, we can use this technique to monitor the electron temperature in the metal film while pulse heating is used to create a nonequilibrium on it. In Fig.~\ref{relaxationmeasurement}(a), we show the electron temperature in the heated metal films in response to a heat pulse applied at various bath temperatures. When the heating amplitude is increased from zero, electron temperature starts to rise and finally reaches the steady state by dissipating the Joule heat through e-p scattering as in steady state experiments. When heating is switched off, electrons start to relax and reach the bath temperature of the refrigerator again. For heat current pulses with inverted polarities, the electron temperatures show identical response as expected.

	Figure.~\ref{relaxationmeasurement}(b) and (c) show the zoom-in of black and red squares of (a) for the bath temperature of 41~mK. In Fig.~\ref{relaxationmeasurement}(b), one can see that electron temperature shows almost linear dependence in time within the first few microseconds of the heating pulse. This can be explained by the weak e-p scattering when the temperature difference between electrons and phonons is small. In this case, almost all heating applied heats up the electrons in the metal film. And electron temperature changes in the metal film can then be written as $\Delta T_e = \frac{P_J\Delta t}{C_e}$. Here $P_J$ is Joule heating power, $C_e$ is heat capacity of the metal film and $\Delta t$ is the time delay between heating and probing pulse. As electron temperature increases further, the contribution of e-p scattering start to dominate and electron temperature shows deviation from linear dependence in time. 
	
	When switching off heating, electrons in metal films start to cool by e-p scattering. For this relaxation process shown in Fig.~\ref{relaxationmeasurement}(c), one can write down the thermal equation 
	\eq{
		C_e\frac{d\Delta T_e}{dt} = -G_{th}\Delta T_e. \label{relaxation}
	}
    From Eq.~(\ref{relaxation}), one finds $\Delta T_e = \Delta T_e(0)e^{-t/\tau}$. The thermal relaxation time is given by $\tau = \frac{C_e}{G_{th}}$. Here, we assume $T_p = T_e$, and that $\Delta T_e(0)$ is small enough so that $C_e$ and $G_{th}$ can be approximated by their equilibrium values. As shown in Fig.~\ref{relaxationmeasurement}(a), the relaxation process is faster at higher bath temperatures. With the technique shown above, one can monitor the electron temperature of metal films in a time-dependent nonequilibrium state. The measurement presented here can be used to determine the heat capacity and thermal relaxation time of normal metal films at sub-kelvin temperatures. The material-specific measurement results will be reported elsewhere.

	The measurement technique for the nonequilibrium experiment with normal metal applies also to other systems once a well-defined Fermi distribution is formed during the measurement. In our experiment, the dimensions of the metal film are sufficiently small and a simulation shows that the whole metal film has a uniform electron temperature while heating is applied. For a system with larger dimensions or with low electron density, a temperature gradient may exist within the system. The measured electron temperature then depends on where the thermometer is located in the system. For a single-shot measurement with JJ thermometer, measurement speed can reach nanosecond range\cite{Zgirski2017}. With the normal metal in-between two superconductors functioning as an absorber, one can use this technique to measure quanta of energy from/to a system, e.g. in form of single-photon detection.

	In conclusion, we have demonstrated the fast thermometry based on proximity Josephson junctions. In a steady state measurement, we determined the heat transport via e-p coupling in Ag films. By employing rectangular pulses for heating and probing, we can monitor the electron temperature in metal films in time-dependent nonequilibrium. The measurement technique presented here can be used to explore phenomena in mesoscopic thermodynamics.
	
	We acknowledge M. Meschke and J. T. Peltonen for technical help, and D. Golubev for useful discussions. This work is supported by the Academy of Finland Center of Excellence program (Project number 284594). We acknowledge Micronova Nanofabrication Centre of Aalto University for providing the processing facilities.

\end{document}